\documentclass[reprint, amsmath,amssymb, aps,prl,superscriptaddress,footinbib]{revtex4-2}

\usepackage{graphicx}
\usepackage{dcolumn} 
\usepackage{xcolor}
\usepackage{bm}
\usepackage{float}
\usepackage{hyperref}
\usepackage{mathrsfs}
\usepackage{siunitx}
\usepackage{ulem}
\newcommand{\ep}{\xi}
\newcommand{\rl}{r_{\ell}}

\renewcommand{\d}{\mathrm{d}}
\newcommand{\lc}{\ell_c}
\newcommand{\lcl}{\ell_c}
\newcommand{\ur}{\mathbf{e_r}}
\newcommand{\ut}{\mathbf{e_{\boldsymbol\theta}}}
\newcommand{\uz}{\mathbf{e_z}}
\newcommand{\uphi}{\boldsymbol{e_\phi}}
\newcommand{\ud}{\mathbf{e_{ij}}}
\newcommand{\lzer}{{\mathscr{L}_i}}
\newcommand{\lun}{{\mathscr{L}_j}}
\renewcommand{\epsilon}{\varepsilon}
\renewcommand{\phi}{\varphi}
\renewcommand{\vec}{\mathbf}

\DeclareMathOperator{\jz}{J_0}
\DeclareMathOperator{\ju}{J_1}
\DeclareMathOperator{\iz}{I_0}
\DeclareMathOperator{\iu}{I_1}

\begin{document}

\title{Orbiting, colliding and merging droplets on a soap film: toward gravitational analogues}

\author{Jean-Paul Martischang}
\affiliation{Univ. Lille, CNRS, Centrale  Lille, Univ. Polytechnique Hauts-de-France, UMR 8520, IEMN, F59000 Lille, France}
\author{Benjamin Reichert}
\email{benjamin.reichert@univ-lille.fr}
\affiliation{Univ. Lille, CNRS, Centrale  Lille, Univ. Polytechnique Hauts-de-France, UMR 8520, IEMN, F59000 Lille, France}
\author{Germain Rousseaux}
\affiliation{Institut Prime (UPR 3346), CNRS - Université de Poitiers - ISAE ENSMA, 11 Boulevard Marie et Pierre Curie Téléport 2 - BP 30179, 86962 Futuroscope Chasseneuil Cedex, France}
\author{Alexis Duchesne}
\affiliation{Univ. Lille, CNRS, Centrale  Lille, Univ. Polytechnique Hauts-de-France, UMR 8520, IEMN, F59000 Lille, France}
\author{Michael Baudoin}
\email{michael.baudoin@univ-lille.fr}
\affiliation{Univ. Lille, CNRS, Centrale  Lille, Univ. Polytechnique Hauts-de-France, UMR 8520, IEMN, F59000 Lille, France}
\affiliation{Institut Universitaire de France, 1 rue Descartes, 75005 Paris}

\date{\today}

\begin{abstract}
Modern telescopes provide breathtaking images of nebulae, clouds and galaxies shaped by gravity-driven interactions between complex bodies.  While such structures are prevalent on an astrophysical scale, they are rarely observed at the human scale. In this letter, we report the observations of the complex orbits, collision, and coalescence of droplets on a soap film, forming structures such as bridges and spiral arms, reminiscent of their astrophysical counterparts. These dynamics emerge from attractive forces caused by gravito-capillary-driven distortions of the supporting soap film. Long orbits and intricate coalescence mechanisms are enabled by the small dissipation in the soap film and the fluidic nature of the droplets and supporting film, respectively. The existence of stable droplets within the soap film featuring a universal radius, as well as the attractive potentials, are explained through a careful comparison of experimental data with models computing the distortions of the supporting soap film. This work opens perspectives to study analogies between phenomena occurring at dramatically different length and time scales. \\

\begin{description}
\item[Keywords]
soap film, liquid lens, potential well, gravitational analogues
\end{description}
\end{abstract}

\maketitle

\begin{widetext}
    \begin{minipage}{\linewidth}
        \begin{figure}[H]
            \centering
            \includegraphics{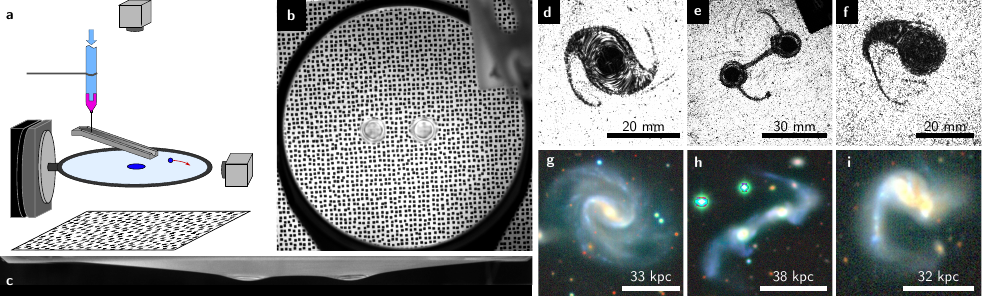}
            \caption{Droplets orbiting and merging on a soap film and resulting structures. \textbf{(a)} Sketch of the experimental setup. \textbf{(b)} and \textbf{(c)} Images of two orbiting drops on a soap film from above (see also Movie M0) and from aside respectively. \textbf{(d-f)} Images of structures resulting from the interactions of two orbiting drops and comparison to similar galaxy structures \textbf{(g-i)}. \textbf{(d)} and \textbf{(e)} feature the interaction of droplets of identical mass (\SI{22.5}{mg} and \SI{18.4} {mg} respectively), whereas \textbf{(f)} illustrates the interaction of droplets of different masses (\SI{38.5}{mg} and \SI{15.5}{mg} respectively). Images \textbf{(g-i)} correspond to the three galaxies  ARP 73, ARP 238 and ARP 55 respectively, taken from the Legacy Surveys / D. Lang (Perimeter Institute)\cite{deyOverviewDESILegacy2019}.}
            \label{fig:experiment}
        \end{figure}    
    \end{minipage}
\end{widetext}

Our universe is shaped by gravitational interactions. Yet the observation of the dynamical organisation of the matter at the astrophysical scales lies beyond the reach of human time-scale observation. This lack of chronological data contributes to some persisting mysteries surrounding the formation and interaction of galaxies \cite{barnesDynamicsInteractingGalaxies1992b, sellwood2022spirals, sellwood2023stability, sun2024role}. 

At the laboratory scale, multiple-body gravity-driven interactions have been widely studied in the field of interfacial fluid dynamics. Indeed, millimetric identical particles deposited over a liquid/air interface naturally experience some gravito-capillary induced interparticle attractive forces resulting from the distortion of the interface \cite{nicolson1949,kralchevsky2000,vella2005}. This effect, commonly referred as the "Cheerios effect", is at the core of capillary self-assembly of micro-objects \cite{cavallaro2011}, used in a wide range of applications ranging from the assembly of microcomponents to the synthesis of functional materials and additive manufacturing \cite{whitesidesSelfAssemblyAllScales2002}. Recently, capillary orbits resembling celestial system have also been reported \cite{gauthier_capillary_2019}. Yet, these simple systems based on a limited number of non-deformable particles cannot give rise to complex structures, such as the ones resulting from the interactions of galaxies.

In this paper, we investigate the orbit, collision, and merging of miscible liquid drops supported by a soap film. We first demonstrate that when a liquid drop is deposited at the center of a soap film, it forms a stable structure with a universal radius, which we refer to here as a "lens". The shape of this lens arises from a balance between gravitational forces and capillary forces induced by the film deformation. Second, we examine the dynamics of a single lens on the fluid membrane, and reveal that the latter acts as a harmonic potential gravito-capillary trap for the suspended droplet. Third, we consider the two-body (lenses) problem on the soap film. The fluidic nature of the system lowers significantly the drag experienced by the droplets, which exhibit sustained orbital motion governed by the background harmonic well on the one hand, and a gravito-capillary pair attraction on the other hand. It appears that the long ranged nature of the the latter effect coupled to droplet deformability allows to evidence physical effects usually observed within the astrophysical realm such as tidal force related deformation and complex merging patterns characteristic of galaxies (see Fig. \ref{fig:experiment}). This work could pave the way towards in-lab study of the dynamics of the formation of gravity driven complex structures observed at the astrophysical scale and hence contribute to the growing field of Analogue Gravity \cite{barcelo2019analogue, rousseaux2020classical, braunstein2023analogue, almeida2023analogue, chernikov_-lab_2012, CRPHYS_2024__25_G1_457_0, viermann2022quantum, banik2022accurate}.

\begin{figure}
    \centering
    \includegraphics{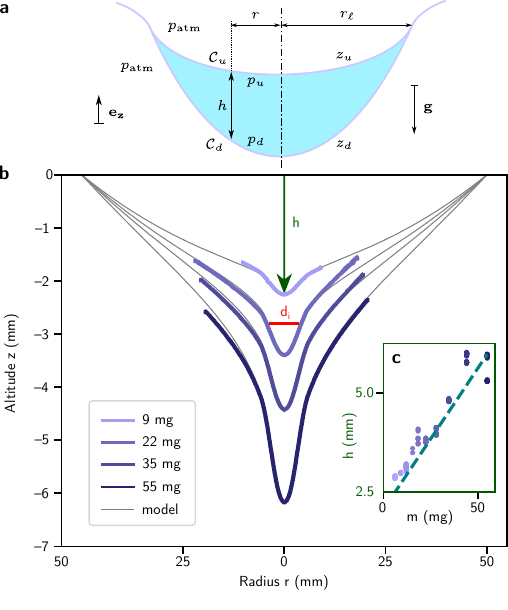}
    \caption{Profiles of a soap film deformed by a water droplet (lens) deposited at its center. \textbf{(a)}, Sketch of the lens and introduction of the parameters of the problem. \textbf{(b)} Superimposed profiles of the bottom interface of the lens captured with a camera for different lens masses (thick purple lines) and comparison with theoretical prediction obtained from equations~(\ref{eq:z}) and~(\ref{eq:zd}) (thin grey lines). The red line with the notation $d_i$ in top indicates the distance between two inflexion points of a given profile. Insert: Evolution of the lowest altitude $h$ of the lens (points) as a function of the droplet mass and comparison to theory (dashed line), corresponding to $h=|z_d(0)|$ in  Eq.~\eqref{eq:zd}.}
    \label{fig:profiles}
\end{figure}   

\paragraph*{\textbf{Experimental setup --}} In the present work, a soap film is formed by immersing a 3D-printed, $10~\si{\centi\metre}$ circular frame into an aqueous surfactant solution (20$\%_v$ -- volumetric percent -- glycerol and SDS at $5~\si{\gram\per\litre}$) and then pulling it out.  The frame, whose shape has been designed to avoid spurious fluid accumulation on the periphery, is maintained horizontal with a goniometer fixed on an anti-vibration table and kept in a sealed box at 80 to 90\% relative humidity to increase the soap film lifetime. Then droplets of water are injected with a syringe either directly on the soap film or using a hydrophobic slide to infer a controlled initial horizontal droplet velocity (Fig.~\ref{fig:experiment}a). The mass of the droplet is adjusted by changing the diameter of the nozzle. Each of the 18 available diameters was thoroughly tested by depositing around 50 droplets on a precision balance, revealing standard deviations in the masses of 2 to 16\% of the observed means, with a median at 5\% of the mean masses. The soap film "in-plane" trajectories of the lenses (Fig.~\ref{fig:experiment}b) and their profile and vertical movements (Fig.~\ref{fig:experiment}c) are recorded  with two orthogonal cameras.

\paragraph*{\textbf{Profile of a single lens --}} When the droplet is deposited at the center of the soap film with a syringe, the liquid remains confined in a central thick region with a "lens" shape, whose weight induces a global deformation of the surrounding thin soap film acting as a fluidic "membrane" (Fig. \ref{fig:profiles}a). This profile evolves slowly, on characteristic time scales \cite{motaghian_rapid_2019} much longer than the present experiments and will hence be considered in the following as quasi-static. The persistence and shape of this lens as well as the global deformation of the soap film membrane can be rationalized by considering the equilibrium between gravity and capillary forces at play:

(i) The equilibrium profile of the fluidic membrane in the peripheral region $r\in[\rl,R]$ can be obtained from a force balance on the soap film, considering the capillary force, the weight of the membrane and the droplet load as a punctual central force of intensity $mg$ (see Supplementary Material (SM) for details):
\begin{equation}
    2\sigma\Delta z(r) = \rho\,g\,\ep + m\,g\,\delta(r),
    \label{eq:equilibrium}
\end{equation} with $r$ the radial distance from the center of the soap film, $\rl$ the radius of the lens, $R$ the radius of the frame, $z$ the vertical displacement of the membrane, $\ep$ the thickness of the soap film considered here as constant, $\sigma$ the surface tension, $m$ the injected droplet mass, $\rho$ the liquid density, $g$ the gravitational acceleration and $\delta$ the Dirac delta function. The resulting profile obtained from the resolution of this linear equation is a superposition of a catenoid induced by the punctual load, and a parabola emerging from the membrane weight:
\begin{equation}
    z(r) =  \frac{m\,g}{4\,\pi\,\sigma}\ln\left(\frac{r}{R}\right) + \frac{\ep}{8\,\lc^2}\,(r^2-R^2).
    \label{eq:z}
\end{equation}

(ii) The upper and bottom interface profiles in the central thick region ($r\in[0,\rl]$) where the drop lies, can be derived from a vertical pressure balance considering Laplace and hydrostatic pressures across the lens:
\begin{equation}
	\sigma\left[\mathcal{C}_u(r) + \mathcal{C}_d(r)\right] = \rho g \,\left[z_u(r)-z_d(r)\right],
    \label{eq:equ_lens}
\end{equation} with $\mathcal{C}_u(r)$ and $\mathcal{C}_d(r)$ the upper and lower curvatures of the lens at distance $r$ from the center of the film and $z_u$ and $z_d$ the upper and lower altitudes of the lens, see Fig.~\ref{fig:profiles}a. The resolution of this equation with a matching of the altitude and slope of the inner and outer solutions for $r=\rl$ leads to the following expressions of the upper and lower interfaces:
\begin{align}
 &  z_u(r) = \frac{\jz(\eta)}{\ju(\eta)}\,\left(\frac{mg}{4\pi\sigma\eta}+\frac{h\eta}{4}\right)\,\left(1-\frac{\iz(r/\lc)}{\iz(\eta)}\right) + z(\rl)    \label{eq:zu} \\ 
& z_d(r) = \frac{\jz(\eta)}{\ju(\eta)}\,\left(\frac{mg}{4\pi\sigma\eta}+\frac{h\eta}{4}\right)\,\left(1-\frac{\jz(r/\lc)}{\jz(\eta)}\right) + z(\rl),
    \label{eq:zd}
\end{align}
with $\eta=\rl/\lc$, $\jz,\ju,\iz$ the Bessel functions of zeroth and first orders and $\lc = \sqrt{\sigma/\rho\,g }$ the capillary length. These results are obtained within the linearized approximation of the curvature.

The composite solution of the bottom interface profile obtained from the combination of Eq.(\ref{eq:zd}) for $r\in[0,\rl]$ and  Eq.(\ref{eq:z}) for $r\in[\rl,R]$ shows excellent quantitative agreement with the one measured experimentally from side view imaging in Fig.~\ref{fig:profiles}b for different droplet masses, with no fitting parameters aside from a unique vertical offset, hence confirming that the physics is well captured by this simple model. The following values for the surface tension and film thickness were considered, measured respectively with the pendant drop method and reflectance spectrophotometry \cite{Lyklema_Scholten_Mysels_1965}: $\sigma=34~\si{\milli\newton\per\meter}$ and $15~\si{\micro\metre}$, the latter corresponding to an average value of the measured thickness spanning between $10~\si{\micro\metre}$ and $20~\si{\micro\metre}$. The experimental maximum depth $h=\big|z_d(0)\big|$ is plotted in insert of Fig.~\ref{fig:profiles}b as a function of the droplet mass, alongside a theoretical prediction (dashed line) obtained by evaluating $\big|z_d(r=0)\big|$ in equation~\eqref{eq:zd}. The predicted linear dependency shows good agreement with the experimental data.

In these equations, the radius $\rl$ is determined through a mass balance $m = \rho \int_{r=0}^{\rl} 2 \pi [z_u(r) - z_d(r)]\,r\, dr$ leading to the implicit equation:
\begin{equation}
    \frac{\jz(\rl/\lc)}{\ju(\rl/\lc)} = -\frac{\iz(\rl/\lc)}{\iu(\rl/\lc)},
    \label{eq:rl}
\end{equation}
A striking feature in this expression (also observed experimentally) is that the lens radial extension $\rl$ ($\approx 6~\si{\milli\metre}$) does not depend on the mass of the droplet.

\begin{figure}
    \centering
    \includegraphics{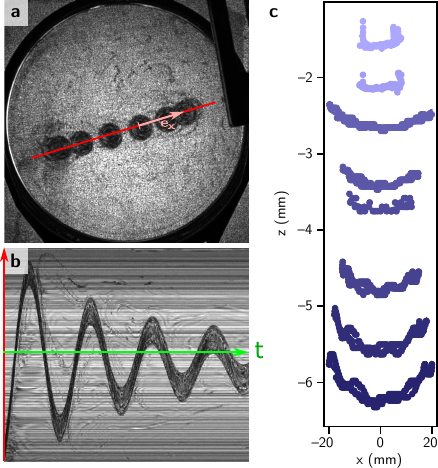}
    \caption{Single droplet moving on a soap film. \textbf{(a)} Superposition of different positions of the droplet. The red solid line shows the axis of of the drop's motion. \textbf{(b)} Temporal evolution of the "in plane" position of the droplet along this line. The droplet motion follows the dynamics of a damped harmonic oscillator. \textbf{(c)} Side view trajectories of lenses of increasing masses (clearer to darker) 6, 9, 28, 31, 35, 38, 51, 55 mg.}
    \label{fig:alone}
\end{figure}

\paragraph*{\textbf{Dynamics of a single off-centered lens --}} Now, if the droplet is injected at an off-center position without initial horizontal velocity, the droplet starts oscillating along an axis $\mathbf{e_x}$ as shown on Fig. \ref{fig:alone}a and b, and Movie M1. Laterally, i.e. in the $(Oxz)$ plane, the droplet trajectory is parabolic (see Fig. \ref{fig:alone}c). The droplet therefore oscillates in a harmonic energy well imposed by the bounded soap film. This trajectory can simply be obtained by generalising Eq. (\ref{eq:equilibrium}) to an off-center position of the lens $\mathbf{r_0}$, i.e. by replacing $z(r)$ by $z(\mathbf{r},\mathbf{r_0})$ and $\delta(r)$ by $\delta(\mathbf{r-r_0})$, with $\mathbf{r}$ the radius vector and $\mathbf{r_0}$ the position of the droplet. Since the droplet oscillates along the axis $(Ox)$, its position is a function only of the coordinates $x_0$ and $z_0$. The relation between these two parameters defining the droplet trajectory can be obtained by solving generalized Eq. (\ref{eq:equilibrium}) (see SM for detailed calculation):
\begin{equation} z_0(x_0) = \frac{1}{8}\frac{\xi}{\lc^2}(x_0^2-R^2) \, + \, \frac{m\,g}{4\pi\sigma} \, \big[\ln(\rl/R)  +(x_0/R)^2   \big].   
\label{eq:trajectoire}
\end{equation}The agreement between this equation and the trajectories observed in Fig.~\ref{fig:alone}c is illustrated in Fig.~\ref{fig:energy}a where the experimental curvature of the lateral trajectory is plotted with respect to the droplet mass, and compared to the trajectory curvature predicted from equation~\eqref{eq:trajectoire}: $A=\frac{1}{8}\frac{\xi}{\lc^2}+\frac{m\,g}{4\pi\sigma R^2}$. Note that this predicted trajectory relies on two underlying assumptions: (i) the membrane is considered at equilibrium and hence the inertial force in the liquid membrane is neglected, and (ii) the inertial force exerted on the droplet in the vertical direction is also neglected.\\

Now that we have determined the lateral droplet trajectory, we can focus on the harmonic oscillatory motion of the drop in the $x$ direction. Projecting Newton's law $m\, \ddot{\mathbf{r_0}} = \mathbf{F_\sigma} +\mathbf{F_{\rm d}} + m\, \mathbf{g}$ onto the base vectors $\mathbf{e_x}$ and $\mathbf{e_z}$ respectively leads to:
\begin{align}
& m \,\ddot{x_0} = \mathbf{F_\sigma}\cdot\mathbf{e_x} + \mathbf{F_{d}}\cdot\mathbf{e_x}, \label{Nx} \\
& 0 = \mathbf{F_\sigma}\cdot\mathbf{e_z} - mg, \label{Nz}
\end{align}
with $\mathbf{F_\sigma}$ the capillary restoring force and $\mathbf{F_d}$ the drag force, which, in the viscous regime, will be proportional and opposite to the velocity vector. This expression is obtained in the low-slope approximation, hence neglecting the speed and acceleration in the $z$ direction. If we now introduce the angle $\beta = \angle (\mathbf{e_z},\mathbf{F_\sigma})$, we obtain straightforwardly $\mathbf{F_\sigma}\cdot\mathbf{e_x} = \tan(\beta) mg$. Hence we just need to determine the angle $\beta$ to identify $\mathbf{F_\sigma}$. This can be achieved (cf. SM) by considering the profile of the deformed membrane for an outcentered droplet, resulting in:
\begin{equation}
    \tan\beta = \frac{\ep x_0}{4\lc^2} + \frac{m\,g}{4\,R\,\pi\sigma} \left[ \frac{x_0/R-x_0^3/R^3}{(1- x_0^2/R^2 )^2 - (x_0\,\rl/R^2 )^2 }\right],
\end{equation}
which is valid in the limit of small droplets compared to the support radius ($\rl \ll R$). The expression for $\mathbf{F_\sigma}$ then reduces to:
 \begin{equation}
  \mathbf{F_\sigma}\cdot\mathbf{e_x} = - \Gamma \,x_0,
\end{equation}
with the supporting film effective spring stiffness
\begin{equation}
    \Gamma = \left[\frac{mg\ep}{4\lc^2} + \frac{(mg)^2}{4\pi\sigma R^2}\right].
\end{equation} 
If we introduce a damping factor $\mathbf{F_{d}}\cdot\mathbf{e_x} = - c \dot{x_0}$ related to the drag (not determined theoretically in the present work), we obtain from Eq.~(\ref{Nx}) the celebrated damped harmonic oscillator equation: 
\begin{equation}
\ddot{x_0} + 2 \alpha \dot{x_0} + \omega_0^2 x_0 = 0,
\label{eq:equa_diff}
\end{equation}
with $\omega_0 = \sqrt{\Gamma/m}$ the natural angular frequency and $\alpha = c/2m$ the damping coefficient. From this calculation, we can introduce a potential energy associated with the droplet lateral movement on the soap film:
\begin{equation}
    E_p = \frac{1}{2}\, \Gamma x_0^2\;.
    \label{eq:ep}
\end{equation} 

\begin{widetext}
    \begin{minipage}{\linewidth}
        \begin{figure}[H]
            \centering
            \includegraphics{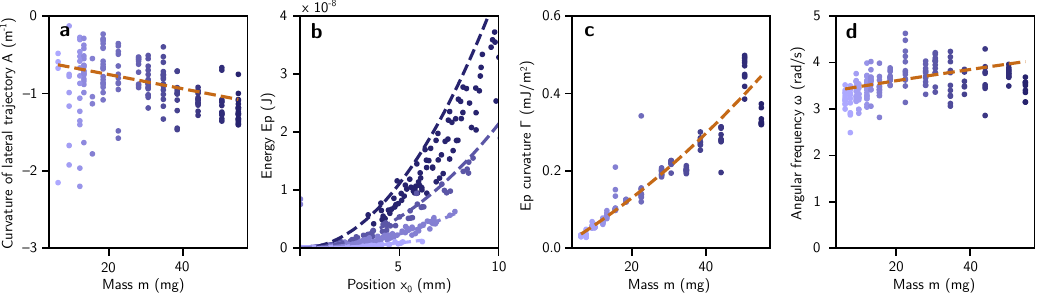}
            \caption{Parameters derived from a lens moving horizontally on a soap film. The experimental points are superimposed with the models developed in this work, in dashed lines. \textbf{(a)} Curvature $A$ of the vertical trajectory $z=A r_0^2$ against the droplet mass $m$. \textbf{(b)} Total energy of the system according to the radial position of the lens for masses 6, 16, 31 and 55 mg (smaller masses are displayed in lighter blue). The $\Gamma$ coefficient of equation~\eqref{eq:ep} is measured from a quadratic fit of the experimental data, and displayed in \textbf{(c)}. The expression of the coefficient $\Gamma$ from equation~\eqref{eq:ep} is indicated in dashed line. \textbf{(d)} Pulsation of the horizontal oscillation of the lens.}
            \label{fig:energy}
        \end{figure}
    \end{minipage}
\end{widetext}

This model can now be compared to the experiments conducted using various droplet masses. First, we take advantage of the damping mechanism to extract the evolution of the potential energy as a function of the radius $x_0$. Indeed, the extremal position reached by the drop $\lvert x_{\rm max}^k \rvert$ during its $k^{\mathrm{th}}$ half-turn on the film decreases at each half period, i.e. for each $k$. By (i) considering  that the potential energy is maximum at these extremal positions of the drop and zero at the center and (ii) neglecting the variation of total mechanical energy (the sum of the kinetic and potential energy) over half a period, we can approximate the potential energy $E_p(\lvert x_{\rm max}^k \rvert)$ as $\frac{1}{2}(E_c^{k-} + E_c^{k+})$, where $E_c^{k-}$ and $E_c^{k+}$ correspond to the kinetic energy of the drop at the the central position preceding and following its passage at the extremum $x_{\rm max}^k$. The results of these measurements are compared to the predictions in Fig.~\ref{fig:energy}b for many different experiments. Then, from these curves, we deduce the value of the effective stiffness $\Gamma$, which corresponds to the curvature of the potential well $E_p(x_0)$ and compare it to the theoretical value (Fig.~\ref{fig:energy}c). Finally, by fitting the damped oscillator curves such as the one represented on Fig.~\ref{fig:alone}b, we determine the damping coefficient (consistently observed around $\alpha=1/3~\si{\per\second}$), as well as the pseudo-pulsation of the oscillator, and compare the latter to its theoretical value $\omega_d = \sqrt{\omega_0^2 - \alpha^2}$ (Fig.~\ref{fig:energy}d). All these figures show quantitative agreement, meaning that the main physics is well captured by our model.

\begin{widetext}
    \begin{minipage}{\linewidth}
        \begin{figure}[H]
            \centering
            \includegraphics{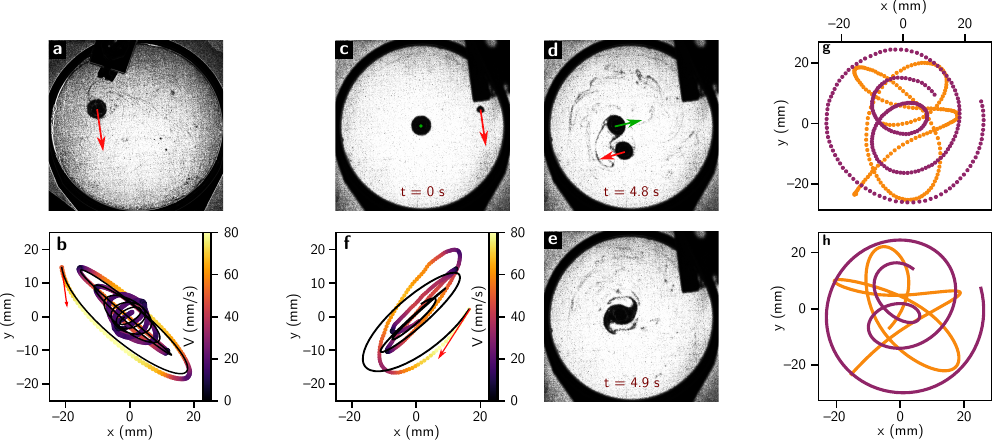}
            \caption{Orbits and merging of lenses. A single 28 mg lens is given an initial horizontal speed (red arrow) in the configuration shown in \textbf{(a)}, resulting the trajectory plotted in \textbf{(b)}, (the color denotes the speed magnitude) and shown in Movie M2. We superimpose this experimental trajectory with a simulated one (black line) based on the model from equation~\eqref{eq:qt} with two oscillators of slightly different pulses $\omega_x$ and $\omega_y$ (the relative difference is about 8\% between the two values). This anisotropy causes a shift in the direction of the orbit during the lens evolution. \textbf{(c)}, \textbf{(d)} and \textbf{(e)} show the evolution of two lenses of 35 mg orbiting and merging in a spiral structure before collapsing to form a new lens (not shown). On \textbf{(c)} and \textbf{(d)}, the green and red arrows indicate the speeds of the lenses (not to scale), with a zero initial speed for the central lens in \textbf{(c)}. The trajectory of the barycenter of this system is plotted in \textbf{(f)} with the same conventions as in \textbf{(b)}, and a black line model for a droplet of mass also 35 mg. \textbf{(g)} Experimental trajectories of both lenses (m=35 mg). \textbf{(h)} Numerical resolution of the dynamics following equation~\eqref{eq:dynamics_two_lenses}.}
            \label{fig:orbits}
        \end{figure}
    \end{minipage}
\end{widetext}

\paragraph*{\textbf{Orbiting droplets --}} Now let's expand the motion of the lens to one additional dimension, by considering its trajectory when it is injected with an initial horizontal velocity using the slide. An example of such trajectory is represented on  Fig.~\ref{fig:orbits}b and Movie M2. To study it, we need to use cartesian coordinates with axes $\mathbf{e_x}$ and $\mathbf{e_y}$, defined along the camera's principal axes for the sake of simplicity and with the center of the soap film as the origin. The potential energy from equation~\eqref{eq:ep} can of course be easily generalized in 2D as: $E_p(x_0,y_0) = \frac{1}{2} \Gamma r_0^2 = \frac{1}{2} \Gamma (x_0^2 + y_0^2)$ since the potential energy only relies on the distance $r_0$ from the center. It can then be used to develop Newton's second law, and project it onto the two orthogonal directions $(Ox)$ and $(Oy)$. It appears that since the potential energy of the system varies quadratically with the distance separating the droplet from the soap film center, the equations of motion in each direction are decoupled and governed by damped harmonic oscillator dynamics. The 2D motion of the droplet is therefore formalised by two orthogonal and decoupled harmonic oscillators whose solution reads:
\begin{equation}
    \begin{cases}
        \displaystyle x_0(t) &\displaystyle = e^{-\alpha t}\,\left[x_i\cos(\omega_x t) + \frac{v_{xi}+\alpha x_i}{\omega_x}\sin(\omega_x t)\right]\\[0.5cm]
        \displaystyle y_0(t) &\displaystyle = e^{-\alpha t}\,\left[y_i\cos(\omega_y t) + \frac{v_{yi}+\alpha y_i}{\omega_y}\sin(\omega_y t)\right].
    \end{cases}
    \label{eq:qt}
\end{equation}
with $x_i$, $y_i$, $v_{xi}$ and $v_{yi}$ the initial coordinate and velocity of the drop along $x$ and $y$ directions respectively and $\omega_x$ and $\omega_y$ the eigenfrequencies in both directions.

\noindent For a perfectly isotropic experiment, the natural frequencies $\omega_x$  and $\omega_y$ would be both equal to $\omega_d$ in equation~\eqref{eq:qt}, leading to a elliptical spiral trajectory toward the center.  However, Fig.~\ref{fig:orbits}b and Movie M2 show that the lens exhibits a complex trajectory progressively drifting toward a straight trajectory before reversing its orbital rotation direction. This behaviour can be rationalized by introducing a slight anisotropy of the soap film leading to  $\omega_x=\omega_d+\delta\omega$ and $\omega_y=\omega_d-\delta\omega$, which induces a progressive drift between the two orthogonal oscillators. By introducing this small anisotropy under the form of $\delta\omega\simeq0.04\,\omega_d$, we see that the model nicely recovers the experimental trajectory, plotted in black on Fig.~\ref{fig:orbits}b. The cause of this anisotropy is yet to be rationalized, but could be explained with inhomogeneities in the soap film thickness or a slight tilt of the frame that prevents perfect horizontality, a well known anisotropic effect of the support for the Foucault pendulum as described first by Kamerlingh Onnes \cite{schulz1970foucault} that induces Lissajoux figures damped here by viscosity.

Now, we can dive into the richest part of this work: the dynamics of two orbiting lenses. A second droplet of water is injected with a non-zero tangential speed around a resting central lens, both having the same masses to begin with. This causes the two bodies to orbit each other for some revolutions and finally to collide and merge (see Fig.~\ref{fig:orbits}c-e and Movie M3). Before the collision, the barycenter of the two drops is expected to follow the same trajectory as a single lens. If we track this barycenter position, and compare it with the previous model with a mass $m$ (the arithmetic mean of the two interacting masses), we indeed obtain a good agreement between our prediction and the experiments, see Fig.~\ref{fig:orbits}f. However, resolving the relative motion of the two drops, as shown in Fig.~\ref{fig:orbits}g, pertains to solving a three-body problem, consisting of each lens and the central force exerted by the supporting film. Hence, it is necessary to add the additional pairwise attraction force between the two lenses. This force derives from an energy of interaction $E_{\rm pair}$ identified as the gravitational potential energy of a given lens  $\lun$ into the deflexion of the film caused by the other one $\lzer)$. Following Nicholson \cite{nicolson1949}, we assume that when two lenses are deposited on the soap film, the total interfacial deformation is the sum of the profiles around the isolated lenses on the soap film (linear superposition approximation). Identifying the interaction pair energy as the product of the weight $m_j\,g$ of $\lun$ with its vertical displacement caused by the profile $z_i$ of the isolated other lens $\lzer$ leads to (details in SM):
\begin{equation}
    E_{\rm pair} = m_j\,g\, z_i(r_{ij}),
\end{equation}
where $r_{ij}$ is the horizontal distance between the two lenses, and the profile of an isolated lens on a weightless membrane $z_i(r_{ij})$ is equal to:
\begin{equation}
    z_{i}(r_{ij}) = \frac{m_i\, g}{4\pi\sigma}\ln\left(\frac{r_{ij}}{R}\right),
\end{equation}
with $m_i$ the mass of $\lzer$.
This simple expression is obtained within the approximation that $\lzer$ and the studied point are both far away from the edge of the film, i.e. when $({r_i\ll R}, {r_{ij}\ll R})$, with $r_i$ the distance separating the center of the soap film from the center of the lens $\lzer$.  As long as these conditions are met -- in fact the approximation appears to stand in experiments even for $r_i$ and $r_{ij}$ being as big as $0.6\,R$ -- we can derive the force driving the second lens $\lun$ towards $\lzer$: ${\mathbf{F}_{\rm pair}=-\bm{\nabla} (E_{\rm pair})}$, i.e.:
\begin{equation}
    \mathbf{F}_{\rm pair} = -\frac{g^2}{4\pi\sigma}\frac{m_i\, m_j}{r_{ij}} \, \mathbf{e_{ij}},
    \label{eq:fpair}
\end{equation}
$\mathbf{e_{ij}}$ being the unit vector between $\lzer$ and $\lun$. Adding this force to equation~\eqref{eq:equa_diff} yields the following dynamics equations driving a lens $i$ pulled by the other lens $j$, in any set of cartesian coordinates $(x,y)$ centered on the film:
\begin{equation}
    \begin{pmatrix}\ddot x_i\\ \ddot y_i\end{pmatrix} + \frac{\Gamma_i}{m_i}\begin{pmatrix}x_i\\ y_i\end{pmatrix}
    + \frac{T_{ij}}{m_i}\cdot\begin{pmatrix}x_i - x_j\\ y_i - y_j\end{pmatrix}
    + 2 \alpha \begin{pmatrix}\dot x_i\\ \dot y_i\end{pmatrix}
    =\begin{pmatrix}0\\ 0\end{pmatrix},
    \label{eq:dynamics_two_lenses}
\end{equation}
where $\Gamma_i$ is the stiffness of the potential well from equation~\eqref{eq:ep} applied to the lens $i$: 
\begin{equation*}
    \Gamma_i=m_i\, g\left(\frac{\ep}{4\lc^2} + \frac{m_ig}{4\pi\sigma R^2}\right),
\end{equation*}and
\begin{equation*}
    T_{ij} = \frac{g^2}{4\pi\sigma}\frac{m_i\,m_j}{r_{ij}^2}\;.
\end{equation*}
Due to the coupling between the dynamics of the two lenses (see equation~\eqref{eq:dynamics_two_lenses}), the resolution of the system dynamics was performed numerically and compared to experimentally trajectories, see Fig.~\ref{fig:orbits}g,h. The results show good agreement with slightly adjusted values of the friction coefficient and film thickness, compared to the values measured during experiments with one lens. It is worth noting that in those simulations, the lenses did go further than $0.4\,R$ from the center, which could also cause some discrepancies between the experiments and simulations.

\begin{figure*}
    \centering
    \includegraphics{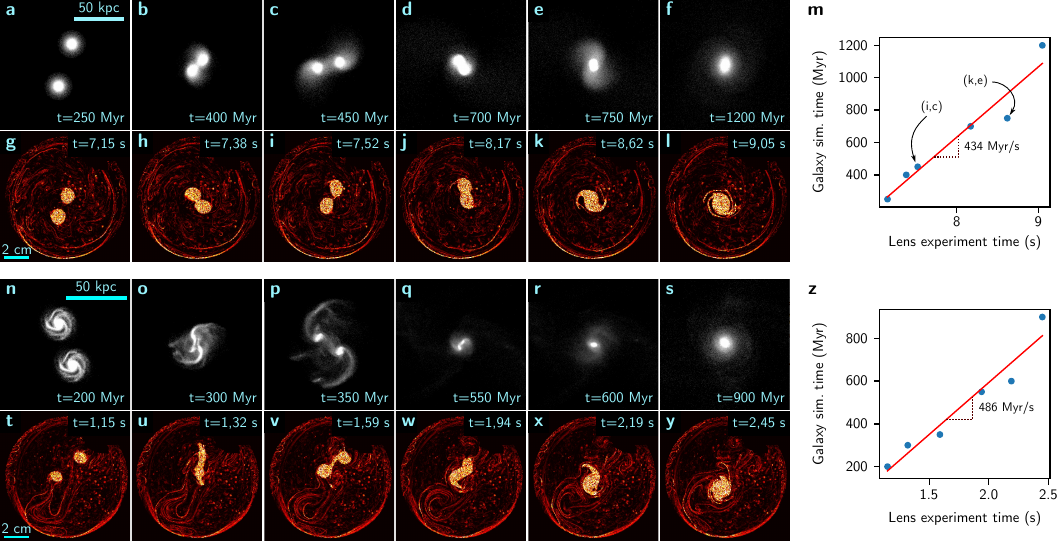}
    \caption{Comparison between simulations of merging galaxies and dynamics of droplets merging on a soap film. \textbf{(a-f)} and \textbf{(n-s)} are sequences of merging galaxies from the GalMer database \cite{chilingarian_galmer_2010}, respectively featuring two giant elliptical galaxies and to giant spiral galaxies. \textbf{(g-l)} and \textbf{(t-y)} are two merging sequences of lenses of same masses (see Movies 5 and 6 respectively), resp. 28 and 18 mg, processed by the Schlieren software ComBOS \cite{mercierEducationalBackgroundOriented2022}. \textbf{(m)} and \textbf{(z)} Correspondence between the timescales of the galaxies simulations (in Myr) and the lens experiments (in s) for each sequence. Each point refers to a couple of images, e.g. i and c, or k and e in \textbf{(m)}.}
    \label{fig:comparaison}
\end{figure*}

\paragraph{\textbf{Merging}} After orbiting, the lenses end up merging into a spiral shape, shown in Fig.~\ref{fig:orbits}e, which, then, collapses to form a new circular lens. Tidal and centrifugal forces are suspected to drive the deformation of the lenses in close proximity of each other, stretching them into the arms of the spiral. For lenses of different masses, we observe that a light body injected with an initial speed will experience far greater deformations than the heavy one it orbits around (see Movie M4). Precise description and modelling of these short-range interactions and deformations, as well as predicting the final spiral shape, are still work in progress and outside the scope of this letter. 

\paragraph{\textbf{Discussion/Conclusion --}}

In this work, we have demonstrated that: (i) a droplet deposited at the center of a soap film induces a deformation (a ‘lens’) with a universal radius independent of the droplet mass; (ii) when a droplet is placed off-center, it experiences a central restoring force  leading to harmonic motion in the absence of initial azimuthal velocity, or complex orbital dynamics otherwise; and (iii) the intricate motion of two droplets orbiting on a soap film is governed by a combination of this harmonic potential well and a pair interaction force $\mathbf{F}_{\rm pair} = -\mathcal{G}_{2D} \frac{m_i\, m_j}{r_{ij}} \, \mathbf{e_{ij}}$, with $\mathcal{G}_{2D} = g^2 / 4 \pi \sigma$, which is essentially a reduced-dimensional (2D) version of Newton’s law of universal gravitation \cite{wilkinsGravitationalFieldsCosmological1986,souchay_tides_2013}.  These effects emerge from the weak deformation of the supporting film, in a manner reminiscent of how Newtonian gravity arises as a weak-field approximation of General Relativity, where gravity is described as the curvature of spacetime. Strikingly, the dynamics and fusion of two identical massive and light drops mirror the simulated mergers of giant elliptical and spiral galaxies respectively (Fig. \ref{fig:comparaison}a-y), albeit at vastly reduced time (and length) scales, with a  scaling correspondence of $\sim 460~\si{Myr\per\second}$. Furthermore, this droplet coalescence sequence produces structures (Fig. 1d-i) that bear a remarkable resemblance to those observed in galaxy mergers.

While this analogy remains in its early stages, the emergence of a Newton-like law of droplet attraction, tidal interactions, and complex merging structures suggests a compelling parallel between these human-scale fluidic experiments and large-scale astrophysical processes, such as galactic fusion. This unexpected bridge between soft matter physics and astrophysics may open new pathways for investigating on a human time scale some simplified representation of phenomena occurring at time scale beyond the reach of humanity.

\acknowledgements \textit{\textbf{Authors contributions.}} M.B. and J.-P.M. first observed the phenomenon, while performing some experiments on extending soap film analogies proposed by G.R. M.B., A.D. J.-P.M. and B.R. designed the experiments. J.P.M.  built the setup and carried out the experiments. M.B., A.D., J.-P.M. and B.R. analysed the data and contributed to the interpretation of the results. B.R. and J.-P.M. derived the theoretical model. M.B., J.P.M. and G.R. proposed and developed the analogy. M.B., A.D. J.-P.M. and B.R. supervised the project. M.B., A.D. J.-P.M., B.R. and G.R. wrote the paper. \\  
\textit{\textbf{Funding:}} Research was funded by IUF, ANR ACOUSURF, ANR-23-CE30-0026 and Université de Lille (project ERC GENERATOR). This work pertains to the French government program ``Investissement d'Avenir'' EUR INTREE (reference ANR-18-EURE-0010) and LABEX INTERACTIFS (reference ANR-11-LABX-0017-01). \\
\textit{\textbf{Conflict of interest:}} The authors report no conflict of interest.

\bibliographystyle{vancouver}

\section{Supplementary material}

\subsection{Equilibrium of a membrane portion}
\begin{figure}[!h]
    \centering
    \includegraphics{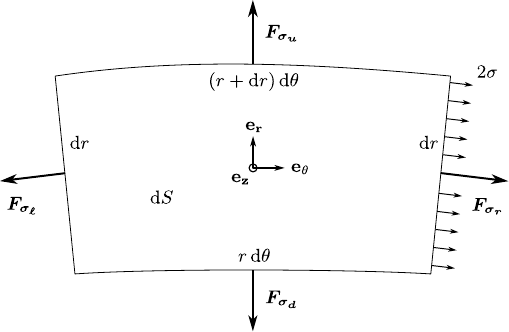}
    \caption{Equilibrium of a portion of the soap film of surface $\d S$, at distance $r$ from the center, over the radius $\d r$ and the angle $\d \theta$. Surface tension $2\sigma$ applies on the four sides of the elementary portion, generating the tension forces $\vec{F}_{\sigma_u}$, $\vec{F}_{\sigma_d}$, $\vec{F}_{\sigma_{\ell}}$ and $\vec{F}_{\sigma_r}$.}
    \label{fig:portion_film}
\end{figure}

The system is parametrized in cylindrical coordinates $(r,\theta,z)$, with the origin $O$ at the center of the film. The profile of the soap film is deduced from the force balance on an elementary portion of area ${\rm dS=r \,dr \,d\theta}$ of the soap membrane (Fig.~\ref{fig:portion_film}). The equilibrium of the fluidic membrane is governed by three forces; its weight denoted $\vec{P} = -\rho\, g\,\ep\,\rm dS\,\uz$, with $\rho$ the density of the liquid, $g$ the gravitational acceleration, and $\ep$ the thickness of the film, a load $-Q(\mathbf{r})\rm dS\,\uz$ of distribution $Q(\mathbf{r})$ hanging on the membrane, and capillary forces exerted on each side of the portion $\rm dS$  of a membrane of tension $2\sigma$,

\begin{equation*}
    \begin{cases}
        \vec{F}_{\sigma_d} &= 2\sigma\cdot r\,\d\theta\;(-\cos\alpha\;\ur - \sin\alpha\;\uz)|_{r}\\
        \vec{F}_{\sigma_u} &= 2\sigma\cdot (r+\d r)\,\d\theta\;(\cos\alpha\;\ur + \sin\alpha\;\uz)|_{r+\d r}\\
        \vec{F}_{\sigma_{\ell}} &= 2\sigma\cdot\d r\;(-\cos\chi\;\ut - \sin\chi\;\uz)|_{\theta}\\
        \vec{F}_{\sigma_r} &= 2\sigma\cdot r\,\d r\;(\cos\chi\;\ut + \sin\chi\;\uz)|_{\theta+\d \theta}\\
    \end{cases}.
\end{equation*}where the inclination of the film portion is parametrized by two angles: $\alpha$ around the axis $\ut$ such as $\tan\alpha=\d z/\d r$, and $\chi$ around $\ur$ with $\tan\chi=\d z/(r\d \theta)$. 

In the small angles approximation ($\sin\alpha\simeq\tan\alpha$, $\sin\chi\simeq\tan\chi$), the projection of the force balance in the vertical direction gives
\begin{equation}
    2\sigma\left[\frac{\partial^2z}{\partial r^2} +\frac{1}{r}\frac{\partial z}{\partial r}+\frac{1}{r^2}\frac{\partial^2z}{\partial\theta^2}\right]\d S - \rho\, g\,\ep\,\d S = 0,
\end{equation}
leading to
\begin{equation}
    2\sigma\Delta z=\rho \, g\, \ep + Q(\mathbf{r}).
\end{equation}

\noindent This Poisson equation shows that the warping of the membrane, of curvature $\Delta z $, is induced by two distinct sources, the membrane weight and the external load of distribution $Q(\mathbf{r})$. In the case of a single droplet hanging on the membrane at location $\mathbf{r}_0$, this load is punctual and its magnitude is identified as the weight $mg$ of the droplet. The load distribution simply becomes  $Q(\mathbf{r})= mg\, \delta(\mathbf{r}- \mathbf{r}_0)$, with $\delta$ being Dirac's delta function. 

\subsection{Equilibrium profile of the soap film}

The equilibrium profile of the membrane for a drop  located at $\mathbf{r}_0=(r_0,\theta_0)$ in cylindrical coordinate, is obtained from the following pressure balance 

\begin{equation}
    2\sigma\Delta z(\vec{r},\vec{r_0}) = \rho\, g \,\ep + m\, g\,\delta(\vec{r}-\vec{r}_0).
    \label{eq_pressure}
\end{equation}

\noindent Since this equation is linear, the equilibrium profile of the membrane $z(\vec{r},\mathbf{r}_0)$ is sought as the superposition of two profiles, each stemming from a contribution in the rhs member of the pressure balance \eqref{eq_pressure}, a parabola $z_{p}(\mathbf{r})$ induced by the membrane weight, and a catenoid $z_c(\mathbf{r},\mathbf{r}_0)$ due to the punctual load: 
\begin{equation}
    z(\vec{r},\vec{r_0}) = z_p(\mathbf{r}) + z_c(\mathbf{r},\mathbf{r}_0).\label{eq_decomp1}
\end{equation}

\noindent The profile of a weighing soap film free from any load is axisymmetric and is obtained from the resolution of 
\begin{equation}
    \Delta z_p = \frac{\rho g\ep}{2\sigma}
\end{equation}as a parabola:
\begin{equation}
    z_p(\vec{r}) = \frac{\ep}{8\lc^2}(r^2-R^2).
    \label{eq:z_p}
\end{equation}

\noindent The profile of a weightless membrane with a droplet positioned at $\mathbf{r}_0$ results from the resolution of the Poisson equation

\begin{equation}
    \Delta z_c = \frac{mg}{2\sigma}\delta(\vec{r}-\vec{r}_0).
    \label{eq:zc_1}
\end{equation}
and is a catenoid:

\begin{equation}z_c(\vec{r},\vec{r_0}) = \frac{mg}{2\sigma}\,G(\vec{r},\vec{r}_0)\label{eq_zc_1}\end{equation}
where $G$ is Green's function for problem \eqref{eq:zc_1} on the disc of radius $R$, satisfying the condition $\Delta G(\vec{r},\vec{r}_0) = {\delta(\vec{r}-\vec{r}_0)}$, with the boundary condition $G(\vec{r},\vec{r}_0)=0$ for $|\mathbf{r}|=R $. In other terms:
 
 \[G(\vec{r},\vec{r}_0)=\frac{1}{4\pi} \ln\left[\frac{r^2+r_0^2-2\,r\,r_0\cos(\theta-\theta_0)}{R^2+r^2r_0^2/R^2-2\,r\,r_0\cos(\theta-\theta_0)}\right]. \]

\noindent The general expression of the equilibrium profile of the soap film eventually writes:

\begin{equation}
\begin{aligned}
z(\mathbf{r}, \mathbf{r}_0) ={} & \frac{\ep}{8\lc^2}(r^2-R^2) \, + \\
      & \frac{mg}{8\pi\sigma}\,\ln\left[\frac{r^2+r_0^2-2\,r\,r_0\cos(\theta-\theta_0)}{R^2+r^2r_0^2/R^2-2\,r\,r_0\cos(\theta-\theta_0)}\right].\label{eq_tot}  
\end{aligned}
\end{equation}


In the case of a droplet lying at the center of the soap film ($\mathbf{r}_0=\mathbf{0}$), the profile of the soap film becomes axisymmetric and simplifies into:

\begin{equation}
    z(r) = \frac{\ep}{8\lc^2}(r^2-R^2) + \frac{mg}{4\pi\sigma}\ln\left(\frac{r}{R}\right) .
    \label{eq:profil_statique}
\end{equation}

\noindent This axisymmetric profile is displayed in Fig.2b in the main text. 

\subsection{Equilibrium profile of the lens}
In this part, the lens is at rest at the center of the film. Due to axisymmetry, the coordinates of any point will be reduced to the radial one, $r$. The equilibrium profile of the thick region of the liquid phase ($r\in[0,\rl]$) is determined in the following. Using the notations defined in Fig.2a in the main text, we first consider two Laplace pressure jumps at the upper and lower interfaces
\begin{equation}
    \begin{cases}
        p_{\mathrm{atm}} - p_u(r) = \sigma\,\mathcal{C}_u(r)\\
        p_d(r)-p_{\mathrm{atm}} = \sigma\,\mathcal{C}_d(r)
    \end{cases}
    \label{eq:laplace_pj}
\end{equation}

\noindent where $\mathcal{C}_u$ and $\mathcal{C}_d$ are the upper and lower interfacial curvatures, and $\sigma$ the surface tension of the liquid phase, supposed equal to that of the soap film. Besides, the hydrostatic pressure balance in the liquid phase gives:

\begin{equation}
    p_d(r) = p_u(r)+\rho\,g\, h(r)
    \label{eq:hydrostatic_pb}
\end{equation}

\noindent with $h(r)$ the thickness of the liquid phase. By combining eq. \ref{eq:laplace_pj} and \ref{eq:hydrostatic_pb}, we can write:
\begin{equation}
    \mathcal{C}_u(r) + \mathcal{C}_d(r) = \frac{h(r)}{\lcl^2},
\end{equation}
where $\lcl^2=\frac{\sigma}{\rho g}$ is the squared capillary length. This writes, in a more detailed fashion:
\begin{equation}
    \frac{\partial^2 z_u}{\partial r^2} + \frac{1}{r}\frac{\partial z_u}{\partial r} - \frac{z_u}{\lcl^2} = -\frac{\partial^2 z_d}{\partial r^2} - \frac{1}{r}\frac{\partial z_d}{\partial r}-\frac{z_d}{\lcl^2}\,.
    \label{eq:equality_profiles}
\end{equation}

\begin{figure}
    \centering
    \includegraphics{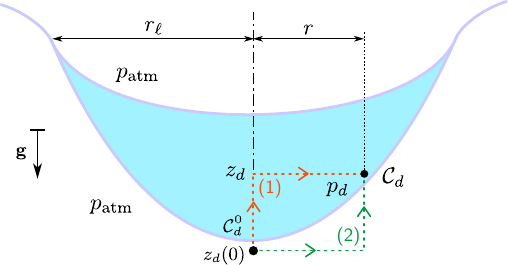}
    \caption{Illustration of two possible path (1) and (2) to evaluate the pressure $p_d(r)$ at the altitude $z_d(r)$, starting from a central point just under the bottom of the film, at $z_d(0)$.}
    \label{fig:paths_pressure}
\end{figure}

Another equation can help us divide this one in two parts. Let's look at the figure \ref{fig:paths_pressure}. Starting from a point just under the lowest point of the film (altitude $z_d(0)$, we can calculate the pressure just above the lower interface for a given radius $r$, $p_d(r)$, either by going up, then right (path 1) or right, then up (path 2) in the convention of the figure. The first case gives us:
\begin{equation}
    p_d(r) = p_{\mathrm{atm}} + \sigma\mathcal{C}_d^0 + \rho g \big[z_d(r)-z_d(0)\big]
\end{equation}
where $\mathcal{C}_d^0$ is the curvature of the lower interface at the center of the film; while the second one yields:
\begin{equation}
    p_d(r) = p_{\mathrm{atm}} + \sigma\mathcal{C}_d(r).
\end{equation}
We can then write:
\begin{equation}
    \mathcal{C}_d^0 -\frac{z_d(r)-z_d(0)}{\lcl^2} = \mathcal{C}_d(r)
\end{equation}
which gives:
\begin{equation}
    \frac{\partial^2z_d}{\partial r^2} + \frac{1}{r}\,\frac{\partial z_d}{\partial r}+\frac{z_d}{\lcl^2} = \mathcal{C}_d^0 + \frac{z_d(0)}{\lcl^2}.
\end{equation}
If we name the constant $C = \mathcal{C}_d^0 + \frac{z_d(0)}{\lcl^2}$, we may split equation \ref{eq:equality_profiles} in two:\begin{equation}
    \begin{cases}
        \displaystyle\frac{\partial^2 z_u}{\partial r^2} + \frac{1}{r}\frac{\partial z_u}{\partial r}-\frac{z_u}{\lcl^2} = -C\\[0.4cm]
        \displaystyle\frac{\partial^2 z_d}{\partial r^2} + \frac{1}{r}\frac{\partial z_d}{\partial r}+\frac{z_d}{\lcl^2} = C\\
    \end{cases}.
\end{equation}
These are two Bessel equations of zeroth order. Avoiding divergence at $r=0$, we get the solutions:
\begin{equation}
    \begin{cases}
        z_u(r)=M \cdot \iz(\frac{r}{\lcl}) + \lcl^2C\\[0.2cm]
        z_d(r)=L \cdot \jz(\frac{r}{\lcl}) + \lcl^2C
    \end{cases}
    \label{eq:zuzd}
\end{equation}
where $M\in\mathbb{R}$, $L\in\mathbb{R}$, and $\jz$ and $\iz$ are Bessel and modified Bessel functions of zeroth order.

The film and the liquid of the lens being miscible, there is no triple line in $r=\rl$, so no reason for a discontinuity in the profile derivatives. We can therefore join $z_u$, $z_d$ and $z$ at $\rl$ with both their values and derivatives. From $\frac{\partial z}{\partial r} = \frac{\partial z_u}{\partial r}$, we get, defining $\eta=\rl/\lcl$:
\begin{equation}
    M = \frac{1}{\iu(\eta)}\cdot\left(\frac{mg}{4\pi\sigma\eta}+\frac{\ep\eta}{4}\right)
    \label{eq:M}
\end{equation}

Then, from $z_u(\rl) = z_d(\rl)$, we get:
\begin{equation}
    L = \frac{\iz(\eta)}{\jz(\eta)} M.
    \label{eq:LM}
\end{equation}

Using this and the fact that $\frac{\partial z_u}{\partial r} = \frac{\partial z_d}{\partial r}$, we can then write:
\begin{equation}
    \frac{\iz(\eta)}{\iu(\eta)} = -\frac{\jz(\eta)}{\ju(\eta)}
    \label{eq:rl_supp}
\end{equation}
with $\ju$ and $\iu$ being Bessel and modified Bessel functions of first order. This gives an expression to numerically determine $\eta$, from which we get $\rl$. Notice that with this, $\rl$ only depends on the capillary length $\lcl$ of the lens, and not on its mass itself.

The last raccording to do is $z(\rl)=z_u(\rl)$, or:
\begin{equation}
    \frac{mg}{4\pi\sigma}\ln\left(\frac{\rl}{R}\right) + \frac{\ep}{8\lc^2}(\rl^2-R^2) = M\cdot \iz(\eta) + \lcl^2C
\end{equation}
from which we get, using equations \ref{eq:M} and \ref{eq:rl_supp}:
\begin{equation}
    \lcl^2C = \frac{\jz(\eta)}{\ju(\eta)}\left(\frac{mg}{4\pi\sigma\eta}+\frac{\ep\eta}{4}\right)+z(\rl).
    \label{eq:lclc}
\end{equation}

After injecting equation~\eqref{eq:lclc} into equations~\eqref{eq:zuzd}, we finally get:
\begin{equation}
    \begin{cases}
        z_u(r) = \frac{\jz(\eta)}{\ju(\eta)}\hspace{-3pt}\left(\frac{mg}{4\pi\sigma\eta}+\frac{\ep\eta}{4}\right)\left(1-\frac{\iz(r/\lcl)}{\iz(\eta)}\right)+z(\rl)\\[0.4cm]
        z_d(r) = \frac{\jz(\eta)}{\ju(\eta)}\hspace{-3pt}\left(\frac{mg}{4\pi\sigma\eta}+\frac{\ep\eta}{4}\right)\left(1-\frac{\jz(r/\lcl)}{\jz(\eta)}\right)+z(\rl)
    \end{cases}.
\end{equation}

\subsection{Trajectory of the lens}
Here, we suppose that the "in plane" -- the plane being $(O,\ur,\ut)$ -- trajectory of the lens is a straight line, confined to a diameter of the soap film, $\theta_0\equiv\rm cst \,[\pi]$. The trajectory of the droplet is obtained from the expression of the membrane equilibrium profile $z(\mathbf{r},\mathbf{r}_0)=z_p(\mathbf{r}) + z_c(\mathbf{r},\mathbf{r}_0) $ (see equations \eqref{eq_decomp1} and \eqref{eq_tot}) evaluated at $\mathbf{r}=\mathbf{r}_0$ ($r=r_0$ and $\theta=\theta_0$) where the droplet center lies. The evaluation of the parabolic contribution to the trajectory is straightforward and gives 

\begin{equation}
    z_{p}(\mathbf{r}=\mathbf{r}_0) = \frac{\ep}{8\lc^2}(r_0^2-R^2).
\end{equation}

The evaluation of the catenoidal contribution to the trajectory is more involved since the catenoid $z_c(\mathbf{r},\mathbf{r}_0)$ diverges at $\mathbf{r}=\mathbf{r}_0$.  The expression of the catenoid only has a physical meaning outside the droplet, i.e. for ${| \mathbf{r}-\mathbf{r}_0 |>\rl}$. Therefore, the vertical position of the center of the lens on this catenoid is defined as the average of the catenoid altitudes on opposite sides of the lens, on $r=r_0-\rl$ and $r=r_0+\rl$:
 
\begin{equation} z_c(\vec{r}=\vec{r_0},\mathbf{r}_0)=<z_c> = \frac{1}{2}[z_c(r_0-\rl)+z_c(r_0+\rl)], \end{equation}
which writes:

\begin{equation}
    <z_c> = \frac{mg}{8\pi\sigma}\Biggl\{2\ln\left(\frac{\rl}{R}\right)-\ln\left[\left(1-\frac{r_0^2}{R^2}\right)^2 - \frac{r_0^2}{R^2}\frac{\rl^2}{R^2}\right]\Biggl\}.
\end{equation}

\noindent In the limit of a small droplet ($\rl\ll R$) oscillating in the vicinity of the center of the soap film ($r_0\ll R$), the altitude of the catenoid at $\mathbf{r}=\mathbf{r}_0$ simplifies into

\begin{equation}
    <z_c> = \frac{mg}{4\pi\sigma}\left[\ln\left(\frac{\rl}{R}\right)+\left(\frac{r_0}{R}\right)^2\right].
\label{eq_alt_catenoid}\end{equation} and the expression of the trajectory $ z_0(r_0)=z(\mathbf{r}_0,\mathbf{r}_0)$ becomes:

\begin{equation}
    z_0(r_0) = \left(\frac{\ep}{8\lc^2} + \frac{mg}{4\pi\sigma R^2}\right) r_0^2 + \frac{mg}{4\pi\sigma}\ln\left(\frac{\rl}{R}\right) - \frac{\ep R^2}{8\lc^2}.
    \label{eq:traj_sup_mat}
\end{equation}

\subsection{Energy of the system}
We aim here to obtain the potential energy from which derives the capillary force attracting the lens towards the center of the film. The motion of the lens in the horizontal plane $(O,\ur,\ut)$ of the soap film is assumed to be a straight line along its diameter.

\begin{figure}[!h]
    \centering
    \includegraphics{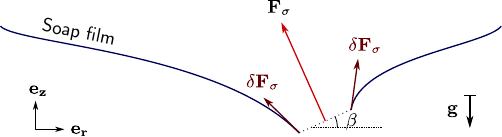}
    \caption{Exaggerated representation of the tilted lens on the soap film, with the capillary forces $\delta\vec{F}_\sigma$ exerted on the contour of the body and their sum $\vec{F}_\sigma$. The lens is tilted with an angle $\beta$ with respect to the horizontal axis.}
    \label{fig:capillary_force}
\end{figure}

In order to derive the capillary force acting on the lens in this plane, we first need to consider a force balance in the plane containing the axis of motion and the vertical direction, namely the $(O,\ur,\uz)$ plane. The two forces acting on the droplet in this plane are the weight of the droplet $-mg \,\uz$ and the surface tension force $\mathbf{F}_{\sigma}=\int_{\mathcal{C}}\delta\boldsymbol{F_\sigma}$, where $||\delta\boldsymbol{F_\sigma}|| = 2\sigma\,\d\ell $, and $\mathcal{C}$ is the droplet contour (see Fig.~\ref{fig:capillary_force}). Since the inertia of the droplet is neglected in the vertical direction, the droplet is in equilibrium in the $z$ direction and the weight of the droplet is balanced by the surface tension force:

\begin{equation*}
    \vec{F}_{\sigma}\cdot\uz - mg = 0
\end{equation*}

\noindent This vertical equilibrium can be recast into an explicit expression by introducing the tilt angle of the droplet $\beta$,

\begin{equation}
    ||\mathbf{F}_{\sigma}|| \cos\beta - mg = 0.
\end{equation}

\noindent The central capillary force attracting any body on the soap film towards its center identifies as the projection of $\mathbf{F}_{\sigma}$ onto the horizontal plane

\begin{equation}
    F_{\sigma r} = \vec{F}_{\sigma}\cdot\ur = -||\mathbf{F}_{\sigma}||\sin\beta=-mg\tan\beta.
\end{equation}The horizontal capillary restoring force is therefore identified as the gravitational force exerted on a droplet in an effective slope $\tan\beta$. Based on the equilibrium shape of the soap membrane, the effective slope has two contributions: the slope due to the parabolic shape of a weighing membrane $\frac{\d z_p}{\d r} (r=r_o)$ and a net slope due to the slope difference of the catenoid on opposite sides of the droplet $<z^{'}_c>= \left[ \frac{\partial z_c}{\partial r}(r=r_0+\rl)   +  \frac{\partial z_c}{\partial r}(r=r_0-\rl)\right]/2$,

\begin{equation}
    \tan\beta = \left.\frac{\mathrm{d}z_p}{\mathrm{d}r}\right|_{r=r_0} +<z^{'}_c>
\end{equation}

\noindent An explicit expression for the net slope is derived from the membrane equilibrium profile (equations~\eqref{eq:z_p} and \eqref{eq_zc_1}):

\begin{equation}
    \tan\beta = \frac{\ep r_0}{4\lc^2} + \frac{m\,g}{4\,R\,\pi\sigma} \left[ \frac{r_o/R-r_o^3/R^3}{(1- r_o^2/R^2 )^2 - (r_o\,\rl/R^2 )^2 }\right].
\end{equation}

\noindent In the limit of a small droplet ($\rl\ll R)$ oscillating in the vicinity of the center of the soap film ($r_0\ll R$), the capillary restoring force writes at first order in $\epsilon=r_0/R$:

\begin{equation}
    F_{\sigma r} = -\left[\frac{mg\ep}{4\lc^2} + \frac{(mg)^2}{4\pi\sigma R^2}\right]r_0
\end{equation}

\noindent The potential energy $E_p$ of this system composed of a droplet hanging on a fluidic membrane is  the potential energy deriving from the capillary restoring force $\rm F_{\sigma r}= -\partial E_p/\partial r_0$, i.e.:

\begin{equation}
    E_p = mg\left(\frac{\ep}{8\lc^2} + \frac{mg}{8\pi\sigma R^2}\right)r_0^2.
\end{equation}We see that the energy of the system varies quadratically with the position $r_0$ of the droplet. It is minimal for a droplet located at the center of the soap film, which therefore acts as a harmonic potential well. 

\subsection{Pair attraction between two lenses}

We consider a soap film loaded with two lenses: one $\lzer$ of mass $m_i$, located at $(r_i,\theta_i)$, and another $\lun$ of mass $m_j$, located at $(r_j,\theta_j)$. We want to determine the pair attraction force between $\lzer$ and $\lun$, via the membrane deformation. The main assumption here is that the total deformation of the interface when two lenses are deposited on the soap film is the sum of the profiles each lens would generate if it was deposited alone. This is the linear superposition approximation Nicholson introduced to derive an analytical expression of the capillary force between two floating bubbles \cite{nicolson1949}.

We start by computing the deformation of the film induced by $\lzer$ (see equation~\eqref{eq_zc_1}), which will be expressed in a new frame centered on $\lzer$. We focus in particular on the deflexion of the membrane along the axis $(\lzer,\lun)$. First, the warping $z_\lzer$
of the membrane at position $(r, \theta)$ caused  by the lens $\lzer$ located at position $(r_i,\theta_i)$ is
\begin{equation}
	z_\lzer(r,\theta) = \frac{m_ig}{8\pi\sigma}\left[2\ln\left(\frac{d}{R}\right) - \ln\left(1-2\epsilon\cos(\theta-\theta_i)+\epsilon^2\right)\right].
\end{equation}with $\epsilon = rr_i/R^2$ and $d = ||\vec{r}-\vec{r}_i|| = \sqrt{r^2+r_i^2 - 2\,r\,r_i\,\cos(\theta-\theta_i)}$ the distance separating our study point from $\lzer$.

Here, we are especially interested in the the warping experienced by the lens $\lun$, i.e. $z_\lzer(r_j,\theta_j)$. If we assume that both lenses are located in the immediate vicinity of the center of the soap film, then $r_i\ll R$, $r_j\ll R$ and $\epsilon\ll 1$. We call $r_{ij}$ the particular value of $d=||\vec{r_j}-\vec{r_i}||$. Then, the first order approximation of the membrane deformation at the lens $\lun$ writes:

\begin{equation}
	z_\lzer(r_j,\theta_j) \simeq \frac{m_i\,g}{4\pi\sigma}\,\left[\ln\left(\frac{r_{ij}}{R}\right) + \epsilon\cos(\theta_j-\theta_i)\right]\,.
	\label{eq:zl_simplifie}
\end{equation}

\begin{figure}[h]
	\centering
	\includegraphics[width=6cm]{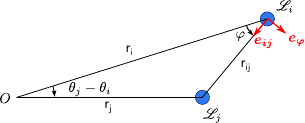}
	\caption{Triangle formed by both lenses $\lzer$ and $\lun$, and the center of the film $O$.}
	\label{fig:triangle}
\end{figure}

In order to recast the membrane deflexion profile (see equation~\eqref{eq:zl_simplifie}) into the frame centered on $\lzer$, the polar angle $\theta$ is replaced by an angle $\phi$ around $(\lzer,\uz)$ and the radial coordinate $r_j$ is replaced by $r_{ij}$ (see Fig.~\ref{fig:triangle}). The vector basis associated to this frame of reference is denoted $(\ud, \uphi)$. This change of variable is performed by applying twice Al-Kashi's theorem on the triangle in Fig.~\ref{fig:triangle}:
\begin{equation}
	\cos(\theta_j-\theta_i) = \frac{r_i - r_{ij}\,\cos\phi}{r_j}.
\end{equation}and allows to determine the profile equation into the new coordinate system $(d,\phi)$:

\begin{equation}
	z_\lzer(r_{ij},\phi) = \frac{m_i\,g}{4\pi\sigma}\left[\ln\left(\frac{r_{ij}}{R}\right) + r_i\,\frac{r_i-r_{ij}\,\cos\phi}{R^2}\right]\,.
\end{equation}

Having calculated the interfacial deflexion caused by $\lzer$,  and knowing the weight of $\lun$, we can now obtain the energy of interaction of the two lenses. This pair energy is the product of the weight of $\lun$ with its vertical displacement along the membrane deformation induced by the presence of its counterpart $\lzer$:

\begin{equation}
E_{\rm pair}= m_j\, g\, z_\lzer
\label{interaction_energy_pair}\end{equation}


Once again, in this analysis, we essentially focus on droplets dynamics occurring in the immediate vicinity of the soap film center which implies $r_i\ll R$ and $r_{ij}\ll R$. At dominant order, the interaction energy depends exclusively on the distance $r_{ij}$ separating the two lenses:

\begin{equation}
E_{\rm pair}= \frac{m_i\, m_j\,g^2}{4\pi\sigma} \ln\left(\frac{r_{ij}}{R}\right)
\label{interaction_energy_pair_simple}\end{equation}

\noindent The expression of the interaction energy reveals that as far as the lenses motion is studied in the vicinity of the center of the soap film, the pair capillary attractive force only depends on the distance separating both lenses, and is directed along the line joining their centers,

\begin{equation}
    \mathbf{F}_{\rm pair} = -\bm{\nabla}(E_{\rm pair})=-\frac{g^2}{4\pi\sigma}\frac{m_i\,m_j}{r_{ij}}\;\ud
\end{equation}where $\ud$ is the unit vector joining the lenses centers (see Fig.~\ref{fig:triangle}).

\end{document}